\documentclass[%
 reprint,
superscriptaddress,
showpacs, 
nofootinbib,
nobibnotes,
 amsmath,amssymb,
 aps,
 prl,
]{revtex4-1}

\usepackage{graphicx}
\usepackage{dcolumn}
\usepackage{bm}
\usepackage{amsmath}
\usepackage{amsfonts}
\usepackage{amssymb}

\begin{document}

\preprint{APS/123-QED}
 
\title{Simultaneous generation and compression of broadband terahertz pulses in aperiodically poled crystals}

\author{Koustuban Ravi}
\email {koust@mit.edu}
\affiliation{%
Research Laboratory of Electronics and the Department of Electrical Engineering and Computer Science, Massachusetts Institute of Technology
}%
\affiliation{
Center for Free-Electron Laser Science, DESY,Notkestra$\beta$e 85, Hamburg 22607, Germany
}%

\author{Franz X. K\"artner}
\affiliation{%
Research Laboratory of Electronics, Massachusetts Institute of Technology
}%
\affiliation{
Center for Free-Electron Laser Science, DESY,Notkestra$\beta$e 85, Hamburg 22607, Germany
}%
\affiliation{
Department of Physics, University of Hamburg, Hamburg 22761, Germany 
}%

\begin{abstract}
We introduce a technique to generate compressed broadband terahertz pulses based on the cascaded difference frequency generation of a non-uniform sequence of pump pulses in aperiodically poled structures. The pump pulse format and poling of crystals conceived are such that the emergent terahertz pulse is already compressed. The approach circumvents pump pulse distortions that result from non-collinear approaches and the need for external compression. It is particularly efficient for the generation of pulses with durations of few to tens of cycles. For instance, calculations accounting for cascading effects predict conversion efficiencies in the few percent range and focused electric fields $\gg$ 100 MV/m in cryogenically cooled lithium niobate. 
\end{abstract}

\maketitle

\section{\label{sec:level1}Introduction}
High-field terahertz radiation are important probes for a variety of fundamental scientific investigations \cite{mishra2013,mishra2015,kampfrath2013}. In addition, they are considered to be invaluable to future particle accelerator technology \cite{palfalvi2014,Nanni2015,RonnyHuang:16,fallahi2016,zhang2018,lemery2018}, which could transform a wide swath of applications encompassing microscopy \cite{eisele2014,cocker2016,ropers2016,kealhofer2016}, X-ray generation \cite{kartner2016} and medical therapy \cite{bravin2012}. 

While various methods to generate terahertz(THz) radiation, ranging from photoconductive switches \cite{jarrahi2015} to Gyrotrons \cite{gold1997review} and free-electron lasers \cite{gallerano2004} exist, they have been limited in one or more of frequency, peak power or bandwidth. With the proliferation of advanced solid-state laser technology, methods based on nonlinear optical frequency conversion of high power infrared or far-infrared radiation have gained ground. Certainly, they have produced the best optical-to-terahertz conversion efficiencies and peak electric fields to date \cite{Vicario:14}.In addition, laser-driven approaches offer precise synchronization possibilities, which are instrumental for spectroscopic experiments.
 
Despite this remarkable process, drawbacks persist with various nonlinear optical frequency conversion schemes. In organic crystals, terahertz and optical radiation are close to being perfectly phase matched, purely by virtue of material properties. Therefore,crystals shorter than the coherence length maybe utilized \cite{Vicario:14,lu2015} to generate broadband terahertz radiation $>$ 1 THz efficiently. However, in general, phase-matching needs to be enforced by developing a suitable mechanism.

A general scheme for broadband phase-matching is via the use of tilted-pulse-fronts (TPF) \cite{hebling2002} in non-collinear geometries. This method, amenable for broadband or single-cycle terahertz pulse generation , has yielded very high conversion efficiencies and pulse energies in the $<$ 1 THz in regime in lithium niobate\cite{hirori2011,fulop2012,fulop2014efficient,huang2015,chen2011}.

However, the approach is ultimately limited by a spatio-temporal break-up of the optical pump pulse upon generating terahertz radiation at high conversion efficiencies \cite{Ravi:14,blanchard2014,Ravi:15,lombosi2015}. This issue may be alleviated  using a superposition of beamlets \citep{Ofori-Okai:16,ravi2016ech} and variants thereof \cite{avetisyan2017,avetisyan2017_2,palfalvi2017}. Alternatively, attempts to eliminate the drawbacks of using TPFs in lithium niobate by utilizing semiconductors as the nonlinear material, which have smaller pulse-front-tilt angles, are being pursued \cite{Polonyi17}. 

Quasi-phase-matching based on periodic inversion of the second order nonlinearity\cite{lee2000,vodopyanov2006,tomita2006,hattori2007,vodopyanov2008,ravi2016} addresses the issues posed by non-collinear phase-matching geometries but is more suited to the generation of narrowband terahertz pulses. Recently, we  suggested the use of aperiodically poled structures \citep{Yahaghi:17} to generate broadband terahertz pulses. Such structures are advantageous in that they eliminate limitations posed by non-collinear phase-matching techniques. They also open up the possibility of re-using the optical pump pulse in subsequent stages. However, they require an external device to compress the chirped terahertz pulses , which increases the complexity of practical implementation. 

In this paper, we introduce an approach which utilizes aperiodic structures to generate broadband terahertz pulses which emerge already compressed. The proposal entails using  chirped crystals in conjunction with a sequence of pulses of non-uniform duration. Such a sequence maybe generated by the superposition of a pair of pulses with different bandwidths, relatively chirped with respect to each other. 

In this method, various terahertz frequencies are phase-matched at different locations in the crystal by virtue of aperiodic poling. At the same time, the relative chirp between the optical pump pulses is set to offset the chirp introduced by aperiodic poling. Therefore, the terahertz frequency $\Omega_1$ generated at location $z_1$ arrives exactly in phase with the terahertz frequency $\Omega_2$ generated at location $z_2$. As a consequence, the terahertz pulse emerges compressed upon exiting the crystal. The collinear geometry addresses the issues of pump pulse distortion suffered by TPF(s) while circumventing the need for external compression demanded by prior proposals of aperiodic structures. 

It is found that the technique presented here is particularly suitable for generating terahertz pulses of few to tens of cycles. As a specific example, the use of cryogenically cooled lithium niobate at 80 K pumped by $1~\mu m$ lasers are predicted to yield optical-to-terahertz energy conversion efficiencies of a few percent with field strengths of 50 MV/m for collimated beams. Similar performance with materials such as Potassium Titanyl Phosphate (KTP) may also be envisaged. The use of chirped $\sim$100 ps pump pulse durations enables significant energy loading of demonstrated cm$^{2}$ aperture poled crystals \cite{ishizuki2014}, resulting in focused field strengths of 500 MV/m and terahertz pulse energies of several mJ. Such high-field, few-cycle pulses could be transformative for terahertz driven electron acceleration, resulting in electron bunches with unprecedented properties which could in turn be potentially disruptive for electron microscopy and X-ray generation.

In subsequent sections, we provide rigorous simulations in both depleted (i.e including cascading effects) and undepleted limits as well as a general analysis to elucidate the physics governing the approach.
\section{\label{sec:2}General physical principle}

\begin{figure*}
\includegraphics[scale=0.5]{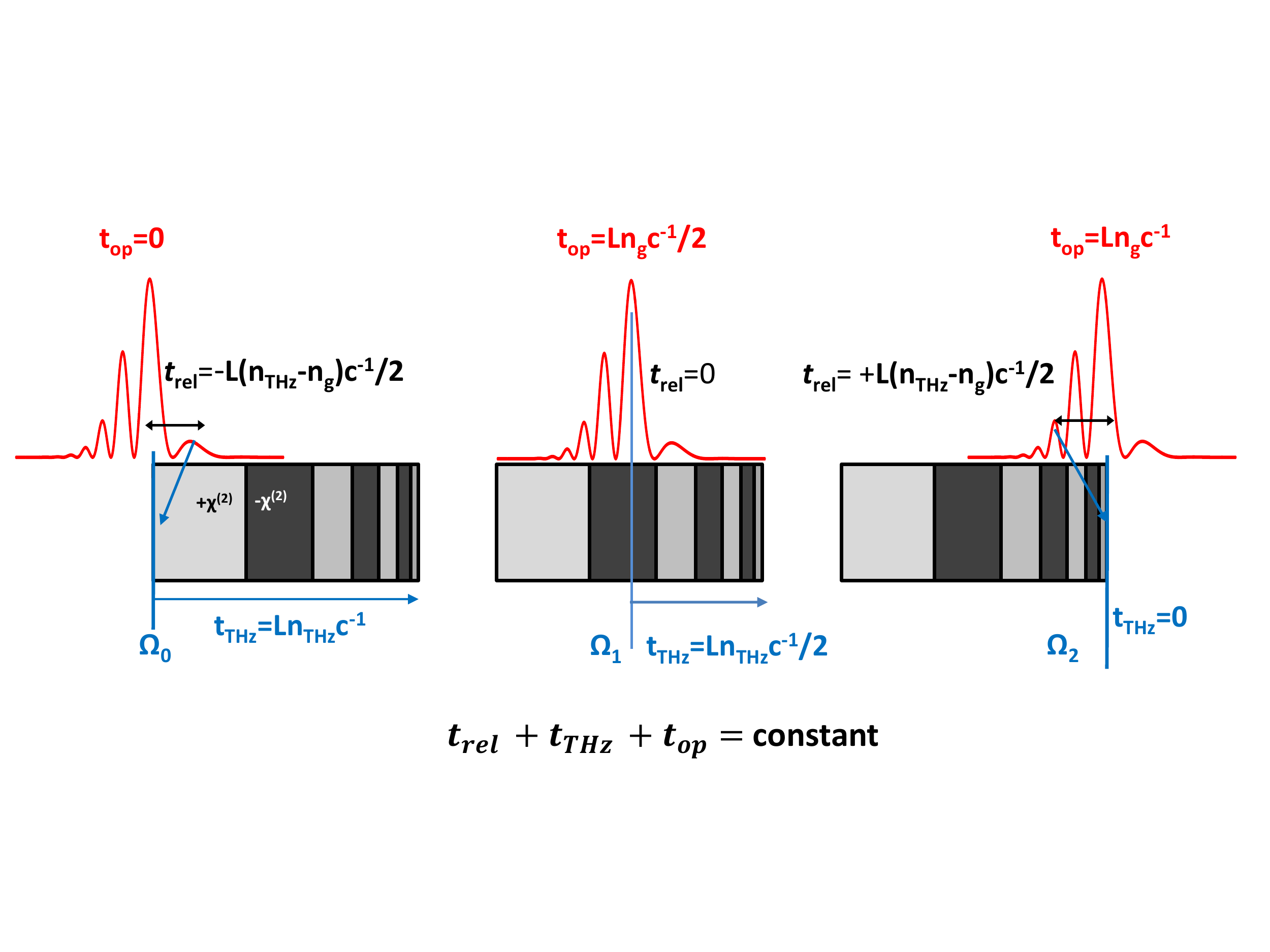}
\caption{\label{fig1}}Schematic of outlined approach : The optical intensity profile appears as a sequence of a  pulses with non-uniform spacing and duration. Lower frequencies are generated at earlier times, $t_{rel}$ in the pulse frame. Simultaneously, the nonlinearity is such that these frequencies are phase-matched earlier in the crystal. The total arrival time $t_{rel}+t_{THz}+t_{op}$ is thus conserved for various terahertz frequencies, thus resulting in compressed terahertz pulses at the output.
\end{figure*}

Consider a situation where terahertz generated at the beginning of a crystal catches up with terahertz radiation generated at the end. This lends itself to a constraint that requires the net arrival time of all terahertz frequency components to be exactly equal. If terahertz radiation of a certain frequency $\Omega_g$ is generated at a certain point $z$ in the crystal , then the following condition conserves the total arrival time at the end of a crystal of length $L$, extending between $z=\mp L/2$:
\begin{gather}\label{eq:1}
\frac{(L/2-z)n_{THz}}{c}+\frac{(z+L/2)n_g}{c}+\Delta t(z)=\textbf{constant}
\end{gather}

In Eq.(\ref{eq:1}), $n_{THz}$ is the terahertz phase refractive index and $n_g$ is the group refractive index of the optical pump pulse. The first term in Eq.\ref{eq:1} corresponds to the time taken by the terahertz frequency component generated at $z$ to arrive at the end of the crystal(i.e. $z=L/2$). The second term in Eq.(\ref{eq:1}) corresponds to the time of arrival of the optical pump pulse at $z$. $\Delta t(z)$ is the relative time delay (in the frame of the pump) at which the given terahertz frequency $\Omega_g(z)$ is generated. For $z=0$,the right hand side of Eq.\ref{eq:1} is $L (n_{THz}+n_g)c^{-1}/2$ as shown in Fig.\ref{fig1}. Consequently, the constraint reduces to a spatially dependent relative delay or $\Delta t(z)$ given by the following :
\begin{gather}\label{eq:2}
\Delta t(z)=\frac{z(n_{THz}-n_g)}{c}=\frac{z\Delta n}{c}
\end{gather}

In principle, the relative delay $\Delta t$ maybe engineered by the use of a sequence of pulses of non-uniform duration as depicted in Fig.\ref{fig1}. Since the locally generated frequency is inversely proportional to the duration of individual pulses in the sequence, the lowest terahertz frequencies are generated at the smallest $\Delta t$, corresponding to the longest pulses in the sequence. Similarly, the largest terahertz frequencies are generated at the longest relative delays , corresponding to the shortest pulses in the sequence.

However, since $\Delta t$ is mapped to each spatial location $z$ by Eq.\ref{eq:2}, we require the lowest terahertz frequencies to be phase-matched at the beginning of the crystal and the largest terahertz frequencies to be phase-matched at the end of the crystal. Thus, the general blue print for producing compressed pulses is established as being feasible by a combination of chirped crystals and a non-uniform sequence of pulses.

\subsection{Practical implementation}
As it turns out, a superposition of (i) a broadband pulse with transform limited duration $\tau_{TL}$ centered at frequency $\omega_1$, and chirped at a rate $b$ to a  duration $\tau$ and (ii) a  narrowband pulse centered at $\omega_2$,  with transform limited duration $\tau$ produce the aforementioned non-uniform sequence of pulses. 

Naturally, the generated terahertz frequency is merely the difference of the instantaneous frequencies of the two pulses given by $\Omega_g=\omega_1+b\Delta t-\omega_2=2\pi f_0+b\Delta t$. However, only the frequency components corresponding to $\Delta t(z)$ from Eq.\ref{eq:2} shall be phase-matched at $z$, which yields the following relationship for terahertz radiation generated as a function of propagation distance $\Omega_g(z)$  :
\begin{gather}\label{eq:3}
\Omega_g(z)=\Omega_0+\frac{bz\Delta n}{c}
\end{gather} 

Equation \ref{eq:3} maybe used to determine a poling profile of the form $\chi^{(2)}(z)=\chi^{(2)}_{b}\textbf{sgn}\left[\text{cos}(\Omega_g(z)(n_{THz}-n_g)zc^{-1})\right]$, where $\chi^{(2)}_b$ is the second-order nonlinear bulk susceptibility and $\textbf{sgn}$ represents the signum function. 

\begin{figure*}
\includegraphics[scale=0.35]{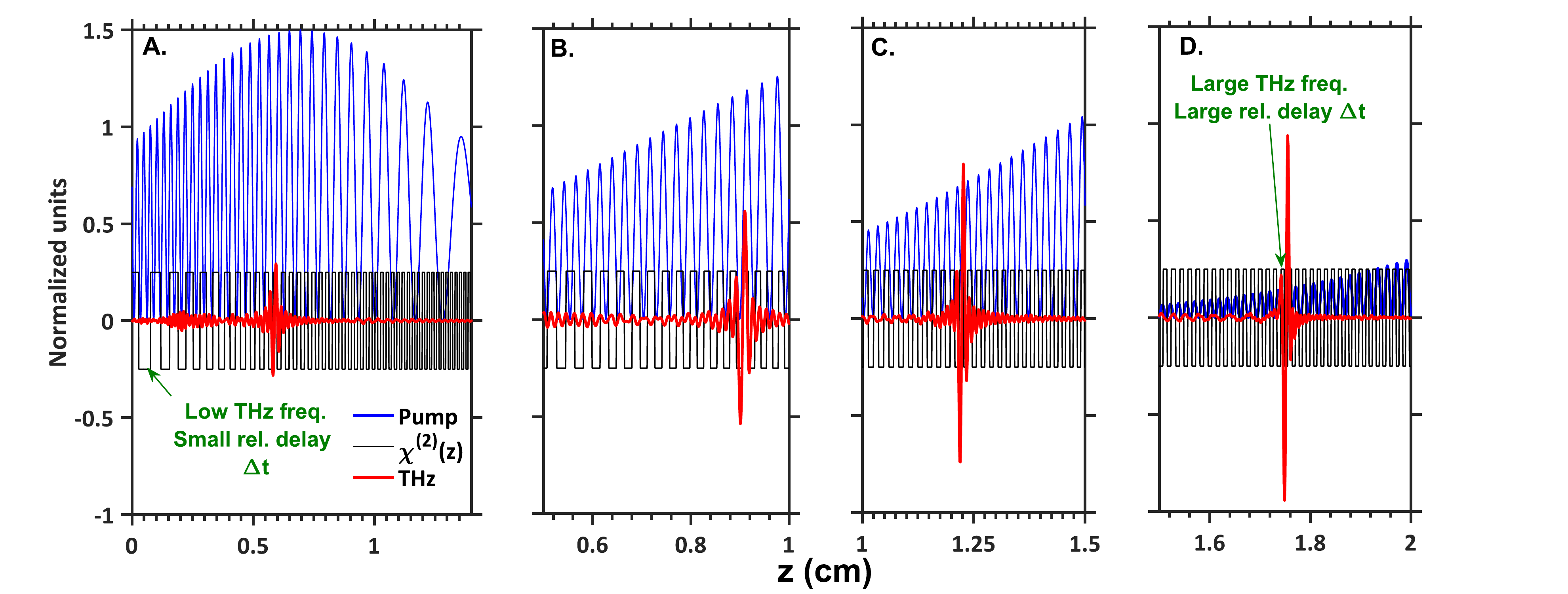}
\caption{\label{fig2}}Numerical simulations depicting a chirped nonlinearity profile along with the intensity profile of the optical pump and terahertz field. The continuous evolution of a compressed terahertz pulse is observed.
\end{figure*}

Numerical simulations employing Eq.\ref{eq:4} for $\tau_{TL}=0.5$ps , $\tau=150$ps and $f_0=0.5$THz in lithium niobate, prove the functionality of the conceived approach as shown in Fig.\ref{fig2}. As can be seen, the optical intensity profile (in blue) is of the form described above. A single-cycle terahertz field (red) with increasing peak field strength, while staying compressed can be seen to evolve.

\section{\label{sec:3}Analytic formulation}
\subsection{\label{subsec:3a}Proof of functionality}
In this section, we will analytically show that the approach proposed in the previous section based on physical reasoning, indeed functions as suggested . We may formally delineate the electric field envelope in time ($t$) of the broadband chirped pulse by the phasor $E_{1}(t)=E_1e^{j\omega_{1}t}e^{jbt^2}e^{-t^2/\tau^2}$ and the narrowband pulse by $E_2(t)=E_2e^{j\omega_{2}t}e^{-t^2/\tau^2}$. This assumption is without loss of generality;the overarching conclusions hold good even for differing pulse durations and chirp rates . As per Eq.(\ref{eq:3}), $2\pi f_0=\Omega_0=\omega_1-\omega_2$. The overall envelope is given by $E(t)=E_1(t)+E_2(t)$. The terahertz spectral component with angular frequency $\Omega$ is given by $E_{THz}(\Omega)=A_{THz}(\Omega)e^{-jk_{THz}(\Omega)z}$, where $A_{THz}(\Omega)=\mathfrak{F}(E(t))$ and $\mathfrak{F}(.)$ corresponds to the Fourier transform between temporal and spectral domains. The corresponding nonlinear evolution of $A_{THz}$ in the undepleted limit is given as follows\cite{Ravi:16}:
\begin{gather}
\frac{dA_{THz}(\Omega,z)}{dz}=\frac{-j\Omega}{2cn_{THz}}\chi^{(2)}(z)e^{j\Omega\Delta nzc^{-1}}\mathfrak{F}\left(|E(t,z)|^2\right)\nonumber\\
-\frac{\alpha(\Omega)}{2}A_{THz}(\Omega,z)\label{eq:4}
\end{gather}
In Eq.(\ref{eq:4}),$\Delta n=n_{THz}-n_g$ represents the difference between the terahertz refractive index $n_{THz}$ at $\Omega_0$ and optical group refractive index $n_g$, $\chi^{(2)}(z)$ is the $z$ dependent second-order nonlinear coefficient and $\alpha(\Omega)$ is the terahertz absorption coefficient. Equation (\ref{eq:4}) maybe reduced to the form below in the absence of material dispersion or loss:
\begin{subequations}
\begin{gather}
A_{THz}(\Omega,L) = \frac{-j\Omega\mathfrak{F}|E(t)|^2\chi^{(2)}_{0}}{2cn_{THz}}\nonumber\\
.\int_{-L/2}^{L/2}e^{-j\Delta kz}\chi^{(2)}(z)dz\label{eq:5a}\\ 
A_{THz}(\Omega,L) = \frac{-j\Omega\mathfrak{F}|E(t)|^2}{2cn_{THz}}\chi^{(2)}_{0}\nonumber\\
.\mathfrak{F}_{\Delta k}\left\lbrace\textbf{rect}\left[\frac{z-L/2}{L}\right]\chi^{(2)}(z)\right\rbrace\label{eq:5b}
\end{gather}
\end{subequations}
The integral in Eq.\ref{eq:5a} may represented as a Fourier transform between the spatial and $\Delta k=-\Delta n\Omega c^{-1}$ domains as shown in Eq.(\ref{eq:5b}). This is possible by casting the situation as one where the nonlinearity $\chi^{(2)}(z)$ is non-zero for  $-L/2<z<L/2$ and vanishes everywhere else in the interval $[-\infty, \infty]$ as evident by the use of the $\textbf{rect}$ function.

One obtains a different perspective when looking at terahertz generation through the lens of Eq.(\ref{eq:5b}). For instance, consider the case of a perfectly periodically poled structure. In such a case, $\chi^{(2)}(z)=\chi^{(2)}_0e^{-j\Omega_0\Delta nc^{-1}}$, where $\chi^{(2)}_0$ is some constant value, proportional to the bulk nonlinearity $ \chi^{(2)}_b$ . In the case of a periodically poled structure, $\chi_0^{(2)} = 2\pi^{-1}\chi^{(2)}_b$. The finite length of the crystal (i.e the $\textbf{rect}$ function in Eq.(\ref{eq:5b})) acts as a filter that produces the all familiar $\textbf{sinc}$ response that characterizes phase-matching in a nonlinear process. 

In the previous section, in Eq.(\ref{eq:3}), the terahertz frequencies $\Omega_g$ generated at various locations in the crystal were deduced for the proposed mechanism.The corresponding poling profile of the form of $\chi^{(2)}(z)=\chi^{(2)}_{b}\textbf{sgn}\left[\text{cos}(\Omega_g\Delta nzc^{-1})\right]$ may be approximated as $\chi^{(2)}(z)=\chi^{(2)}_0e^{-j\Omega_g\Delta nc^{-1}z}$, wherein $\Omega_g$ is given by Eq.(\ref{eq:3}).  Furthermore, as a limiting case, the impact of the \textbf{rect} function in Eq.(\ref{eq:5b}) maybe omitted, thereby resulting in $A_{THz}$ being directly proportional to the Fourier transform of the nonlinearity profile. This approximation may be justified by the fact that due to the finite bandwidth of the input pump pulses, the contributions of the crystal beyond the interval $[-L/2,L/2]$ are less significant. Upon  explicit evaluation of $\mathfrak{F}(|E(t)|^2)$, we obtain the generated spectral component $A_{THz}(\Omega,L)$ in Eq.(\ref{eq:7a}) below.
\begin{gather}
A_{THz}(\Omega,L/2)=\frac{\Omega\chi^{(2)}_{0}E_1E_2e^{-\frac{(\Omega-\Omega_0)\tau_{TL}^{2}}{2(4\tau_{TL}^2\tau^{-2}+1)}}}{4b\Delta nn_{THz}}\nonumber\\
.\left(e^{-\frac{jb(\Omega-\Omega_0)^2}{4(b^2+4\tau^{-4})}}e^{\frac{j(\Omega-\Omega_0)^2}{4b}}\right)\label{eq:7a}
\end{gather}
Inspecting the first exponential term in Eq.(\ref{eq:7a}), we notice that the terahertz spectrum $A_{THz}(\Omega,L)$ is centered about the angular frequency $\Omega_0$ with bandwidth proportional to $(8\tau^{-2}+2\tau_{TL}^{-2})^{1/2}$. In the limit that the chirped pulse duration is much greater than the transform limited bandwidth, i.e. $\tau >>\tau_{TL}$, the terahertz spectral bandwidth is proportional to that of the broadband pump pulse $\tau_{TL}^{-1}$. This corresponds to the range of possible beat frequencies between the two input pulses and is consistent with expectations. The terms enclosed within brackets in Eq.(\ref{eq:7a})correspond to the various phase terms that accrue. The first phase term arises from the difference frequency generation process, while the second term arises from the nonlinear chirped profile. Notice that the two terms are of opposite signs, indicative of a phase compensation effect. 

Given that the chirp rate  $b =(\tau \tau_{TL})^{-1}$ and that the chirped pulse duration is significantly larger than the transform limited duration , i.e. $\tau >> \tau_{TL}$, the factor $e^{\frac{-jb(\Omega-\Omega_0)^{2}}{4(b^{2}+4\tau^{-4})}}$ is well compensated by the $e^{\frac{jb(\Omega-\Omega_0)^2}{4b^{2}}}$ term. In fact, one may achieve exact compression by adjusting the chirp rate of the crystal to $b'=b\sqrt{1+4\tau^{-4}b^{-2}}$.  Thus, it has been analytically proven that the proposed approach produces a perfectly compressed terahertz pulse in the ideal limit. 

The qualitative trends deduced above are supported by calculations employing closed form expressions based on imaginary error functions  (Eq.\ref{eq:8a}), numerical simulations in undepleted (Eq.\ref{eq:4}) and depleted limits (See Fig.\ref{fig7}).

\subsection{\label{sec:3b}Length and frequency behavior}
In Eqs.(\ref{eq:8a})-(\ref{eq:8c}) , we present an analytic solution to Eq.(\ref{eq:4}). In contrast to complete numerical simulations, these expressions only account for the principal harmonic components of the aperiodic poling profile, i.e. $\chi^{(2)}(z)\approx \chi^{(2)}_0e^{-j\Omega_g\Delta nc^{-1}z}$ in the expressions below. 

\begin{subequations}
\begin{gather}
A_{THz}(\Omega,z)=\frac{\Omega\chi^{(2)}_{0}E_1E_2e^{-\frac{(\Omega-\Omega_0)\tau_{TL}^{2}}{2(4\tau_{TL}^2\tau^{-2}+1)}}}{8\sqrt{bb'}\Delta nn_{THz}}\nonumber\\
.\left(e^{-\frac{jb(\Omega-\Omega_0)^2}{4(b^2+4\tau^{-4})}}e^{\frac{j(\Omega-\Omega_0)^2}{4b'}}\right)\textbf{G}(\gamma,\theta)\label{eq:8a}\\
\textbf{G}(\gamma,\theta)= e^{-\gamma\theta/2}\bigg(\textbf{erfi}\bigg[\frac{\sqrt{j}}{2}\big(j\gamma-
\theta\big)\bigg]\nonumber\\
-\textbf{erfi}\bigg[\frac{\sqrt{j}}{2}\big(j\gamma_{z=-\frac{L}{2}}-\theta\big)\bigg]\bigg)\label{eq:8b}\\
\gamma = \frac{2z}{z_0}-\frac{\Omega-\Omega_0}{b'^{1/2}}, ~\theta = \frac{\alpha z_0}{2},~z_0 = \frac{c}{b'^{1/2}\Delta n}\label{eq:8c}
\end{gather} 
\end{subequations}

In Eq.\ref{eq:8a}, the phase terms within brackets correspond to those from the chirped pulses and crystal respectively. As shown in the previous section, these tend to compensate each other. The dependence of the terahertz spectral component $A_{THz}$ on length $z$, terahertz frequency detuning  $\Omega-\Omega_0$ from the central frequency $\Omega_0$ and terahertz absorption coefficient $\alpha$ are captured by the function $\textbf{G}(\gamma, \theta)$ which is delineated in Eq.(\ref{eq:8b}). The variables $\gamma,\theta$ are normalized variables which capture various dependencies as depicted in Eq.(\ref{eq:8c}).In, Eq.(\ref{eq:8b}), the pre-factor $e^{-\gamma\theta/2}$ corresponds to the loss terms contained in Eq.(\ref{eq:7a}). Note that $\gamma$ captures propagation and bandwidth effects, while $\theta$ predominantly relates to the effect of absorption. An important physical quantity here is the distance parameter $z_0= cb'^{-1/2}\Delta n^{-1}$. Since the chirp rate $b'\approx(\tau\tau_{TL})^{-1}$, $z_0= c\Delta n^{-1}\sqrt{\tau\tau_{TL}}$. It may be therefore viewed as some sort of cut-off interaction length beyond which terahertz generation becomes significant. 

To deduce general trends from Eq.\ref{eq:8a}-\ref{eq:8c}, we eliminate dependence on the $L/2$ parameter present in the second term within the brackets of Eq.\ref{eq:8a}, by taking the limit of $-L/2\rightarrow-\infty$ in Eq.\ref{eq:gplot}. This approximation is justified by the fact that the pump pulse contains a finite bandwidth. Therefore local poling periods outside the pump bandwidth will not affect the results or qualitative trends significantly.

\begin{gather}
\textbf{G}(\gamma,z)= e^{-\gamma\theta/2}\bigg(\textbf{erfi}\bigg[\frac{\sqrt{j}}{2}\big(j\gamma-
\theta\big)\bigg]+j\bigg)\label{eq:gplot}
\end{gather}

\begin{center}
\begin{figure}
\includegraphics[scale=0.5]{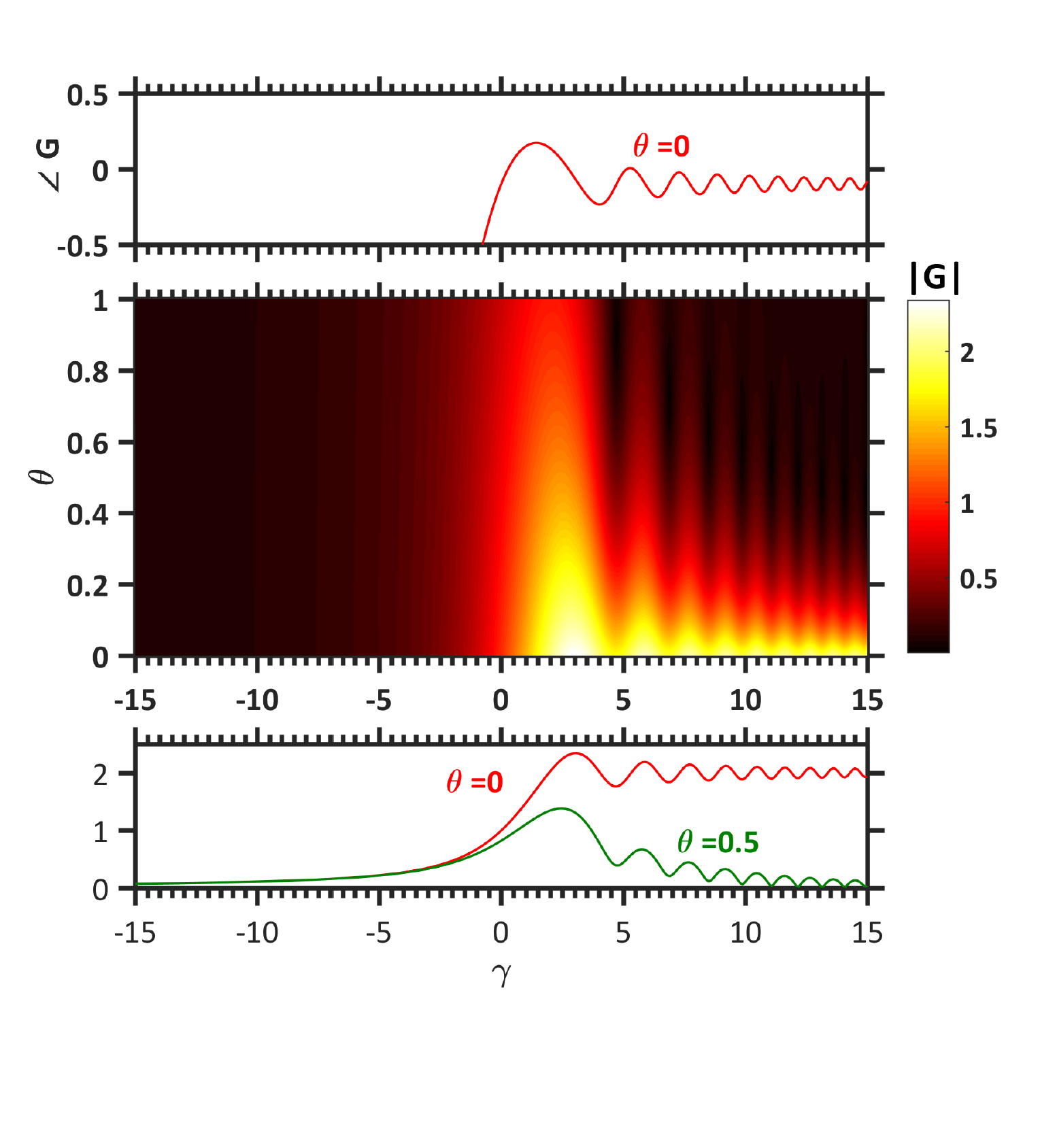}
\caption{\label{fig3}} Plotting the magnitude and phase of \ref{eq:gplot} as a function of the normalized variables $\gamma$ and $\theta$ which simultaneously account for the combined effects of bandwidth, propagation length, $\tau_{TL},\tau$ and absorption coefficient $\alpha$. Since $\angle G\approx 0$, the compression effect is evident.
\end{figure}
\end{center}

Plotting Eq.\ref{eq:gplot} as a function of$\gamma,\theta$ in Fig.\ref{fig3} allows us to discern the general behavior of such systems . The surface plot shows the variation of the magnitude $|\textbf{G}|$. The  panel on top shows the phase $\angle G$ , while the panel at the bottom represents slices of the surface plot as a function of $\gamma$ at various $\theta$ (or absorption) values. The behavior of $\textbf{G}$ is step-like  with its magnitude abruptly increasing before saturating. Naturally, with increasing $\theta$ , an exponential drop in $|\textbf{G}|$ due to absorption effects is evident. 

Since $\gamma$ is proportional to $z/z_0$ and since $|\textbf{G}|$ saturates on average beyond a critical value of $\gamma$, it implies that $z_0$ is a critical length or 'interaction length'  parameter. Both magnitude and phase behavior show rapid oscillations for smaller $\gamma$, but an eventual damping of oscillations for larger $\gamma$. The stabilization of phase over the scale of the critical propagation distance  $z_0$ indicate that the compression is not instantaneous but occurs over a certain distance. Since $|G|$ varies with frequency and absorption coefficient, ripples in the spectral magnitude maybe expected. Furthermore, oscillations in spectral phase about zero are also evident from the behavior of $\angle G$. For the same $z$, $\gamma$ values decrease with $\Omega$.  Therefore, in conjunction with the pre-factor $\Omega$ in Eq.(\ref{eq:8a}), the spectral amplitude would show a ramp like shape, i.e. increasing with frequency as can be seen in Fig.\ref{fig5}a. However, the falling edge of the spectrum shall be also influenced by the exponential dependence on $e^{-2(\Omega-\Omega_0)^2\tau_{TL}^2}$ in Eq.(\ref{eq:8a}). Finally, since $z_0$ decreases with both $\tau,\tau_{TL}$, the optimal interaction lengths shall be expected to be shorter for smaller $\tau,\tau_{TL}$. These deductions are supported by numerical simulations in the next section (See Fig.\ref{fig4}(d)).

\subsection{\label{sec:3c}Efficiency trends}
The upper limit of Eq.(\ref{eq:8a}), in the undepleted limit, is given by Eq.(\ref{eq:7a}). Some general trends on conversion efficiency can thus be gleaned by utilizing Eq.({\ref{eq:7a}). The conversion efficiency $\eta$ may be expressed as follows:
\begin{subequations}
\begin{gather}
\eta = \pi c\varepsilon_0\int_0^{\infty}n_{THz}|A_{THz}(\Omega)|^2d\Omega/F_{pump}\label{eq:9a}\\
\eta \leq \frac{\pi\chi_b^2F_{pump}\Omega_0^2\tau_{TL}}{2\sqrt{2}c\varepsilon_0 n_{THz}\Delta n^2n_{g}^2}\label{eq:9b}
\end{gather}
\end{subequations}
Based on Eq.(\ref{eq:9b}), the proportionality to $\tau_{TL}$ delineates that the conversion efficiency is inversely proportional to the generated terahertz bandwidth. This is understandable given that a larger bandwidth, when apportioned over the same length, translates to less interaction length per frequency. Thus, it is presumable that the conversion efficiency that is generated from such structures would be lesser than that of multi-cycle generation methods. On the other hand, it can be seen that despite this limitation, the conversion efficiency for moderate terahertz pulse durations breaches the percent level. Furthermore, the conversion efficiency is inversely proportional to the difference in terahertz phase and optical group refractive index $\Delta n$. In light of this, KTP with smaller values of $\Delta n$ and potentially larger damage fluence thresholds may offer similar if not superior performance in relation to lithium niobate.

\section{\label{sec:4}Simulation results}
In this section, we present simulation results incorporating the full effects of dispersion, absorption, exact spatial variation of $\chi^{(2)}$ as well as pump depletion. We first obtain optimal parameters via a numerical solution of Eq.\ref{eq:4} and  then verify functionality of the approach in the presence of cascading effects.  While the simulation results are carried out for lithium niobate, the illustrated trends apply to other material systems. The material parameters and numerical model are the same as those utilized in \cite{ravi2016}. The total pump fluence that may be impinged on the crystals, is assumed to be limited by damage according to the following expression for fluence, $F_{\text{pump}} =  0.85(\tau/100\text{ps})^{0.5} Jcm^{-2}$, where $\tau$ is the $e^{-2}$ pulse duration.
 
\subsection{\label{sec:4a}Efficiency and peak fields}
\begin{figure}
\includegraphics[scale=0.4]{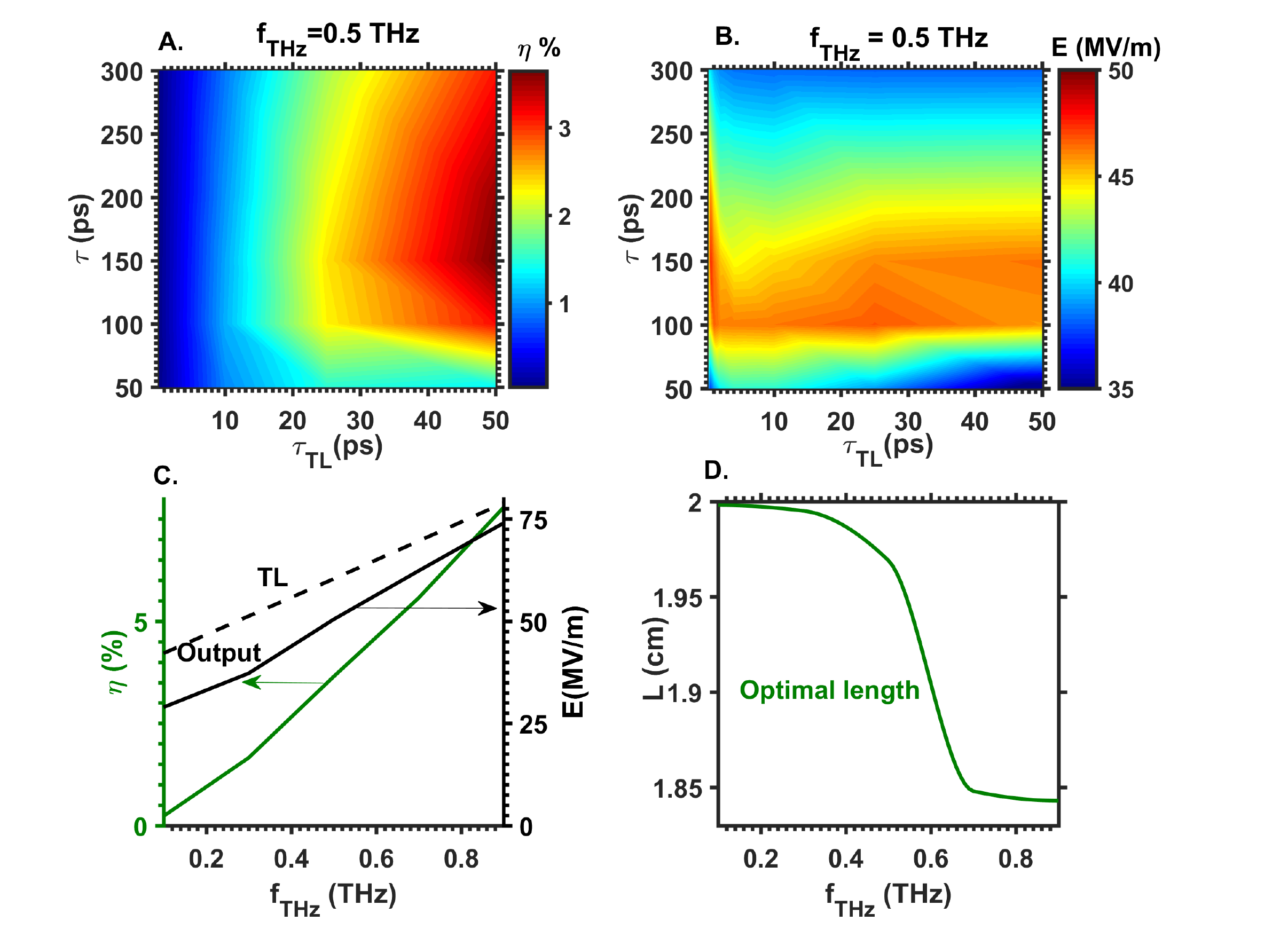}
\caption{\label{fig4}(a)Conversion efficiency $\eta$ as a function of chirped pump pulse duration $\tau$ and transform limited pump duration $\tau_{TL}$ for 0.5 THz in lithium niobate at T=80K. Increasing $\eta$ for narrower bandwidths is predicted, reaching about $4\%$ for $\tau_{TL}$=50 ps. (b) Corresponding electric field strengths in free-space as a function of $\tau,\tau_{TL}$, showing collimated field strengths on the order of 50 MV/m. (c) Optimal $\eta$ and peak electric fields as a function of terahertz frequency, depicting an increasing trend and proximity of the peak fields to the corresponding transform limited values. The optimal values occur for $\tau_{TL}$=50 ps and $\tau\approx 2\Delta nc^{-1}\alpha^{-1}\approx 150$ ps (d)Optimal interaction lengths decrease with frequency, in agreement with increasing absorption coefficients with terahertz frequency.} 
\end{figure}
In Fig.\ref{fig4}, we study the dependence of the optical-to-terahertz energy conversion efficiency ($\eta$), the electric field (E) as well as optimal crystal lengths as a function of the transform limited bandwidth of the pump pulse $\tau_{TL}$, the chirped pulse duration $\tau$ (or equivalently the chirp rate $b$) and the terahertz frequency $f_{THz}$ using undepleted calculations. The bandwidth of the generated terahertz pulse, is limited by $\tau_{TL}^{-1}$, or the range of beat frequencies between the two optical pump pulses. Since, the crystal is aperiodic, each terahertz frequency is generated at a designated region within the crystal. Therefore, the bandwidth of the generated terahertz pulse is also correlated to the chirp rate (w.r.t. length) as shown in Eq.\ref{eq:2} or $b\Delta nc^{-1}L_{opt}\approx \tau_{TL}^{-1}$. However, since $b= (\tau\tau_{TL})^{-1}$, one obtains a condition for the optimal crystal length $L_{opt}\approx c\tau\Delta n^{-1}$, for a given chirped pulse duration $\tau$.

 However, the maximum interaction length is in turn limited by the terahertz absorption coefficient. Thus, $L_{opt}\approx 2\alpha^{-1}$ and consequently $\tau_{opt}\approx 2\alpha^{-1}\Delta nc^{-1}$. As the bandwidth increases, the maximum absorption length reduces by virtue of spanning larger frequencies and hence both the optimal chirped pulse duration and crystal length drop. These aforementioned trends are evident in Fig.\ref{fig4}(a) which depicts $\eta$ as a function of $\tau$ and $\tau_{TL}$ for $f_{THz}=$0.5 THz. Furthermore, in line with the efficiency relationship with $\tau_{TL}$ delineated in Eq.\ref{eq:9b}, we see an increase in efficiency with increasing $\tau_{TL}$.  Intuitively, this may be understood by the fact that when the same length is apportioned over a larger bandwidth, the coherent build up of each terahertz spectral component is limited. Therefore for single-cycle pulses, the efficiency is limited to a fraction of a percent but for few-cycle or tens of cycles long terahertz pulses, the conversion efficiencies breach the percent mark comfortably. Note however that contrary to Eq.\ref{eq:9b}, the increase in efficiency with $\tau_{TL}$ in Fig.\ref{fig4}(a) is not linear as Eq.\ref{eq:9b} represents the ideal limit.
 
In Fig.\ref{fig4}(b), the peak electric fields as a function of $\tau,\tau_{TL}$ are plotted. Since the emergent terahertz pulses are compressed, the peak fields are proportional to $\eta/\tau_{TL}$. Therefore, for small $\tau_{TL}$, the peak electric fields are large despite the low conversion efficiencies. After a minor drop for intermediate values of $\tau_{TL}$, the peak fields tend to stabilize or saturate.The maximum field strengths for collimated beams in free space are on the order of $50~ \text{MV/m}$. However, if one uses cm$^2$ aperture crystals, focused field strengths of several hundred MV/m are indeed feasible.
 
Figures \ref{fig4}(c) and(d) delineate the global optima (w.r.t to parameters $\tau, \tau_{TL}$ for conversion efficiencies and fields obtained for various central terahertz frequencies as well the corresponding optimal crystal lengths. Naturally, these represent the many-cycle limit, i.e. $\tau_{TL}=\text{50~ps}$ or the edge of the parameter space used for $\tau_{TL}$. The corresponding optimal values of $\tau\approx2\alpha^{-1}\Delta nc^{-1}\approx\text{150~ps}$. In Fig.\ref{fig4}(c), the conversion efficiency shows an increasing trend with frequency, albeit only linearly and not quadratically owing to the effects of increasing absorption with terahertz frequency. The maximum electric fields also increase with frequency and are very close to the transform limited values, further depicting the efficacy of the presented mechanism in achieving compression. Figure \ref{fig4}(d) depicts the optimal interaction lengths as a function of frequency. Here, too the trends of decreasing optimal lengths are consistent with increasing absorption coefficients with terahertz frequency.

\subsection{\label{sec:4b}Spectra, phase and temporal formats}
\begin{figure*}
\includegraphics[scale=0.5]{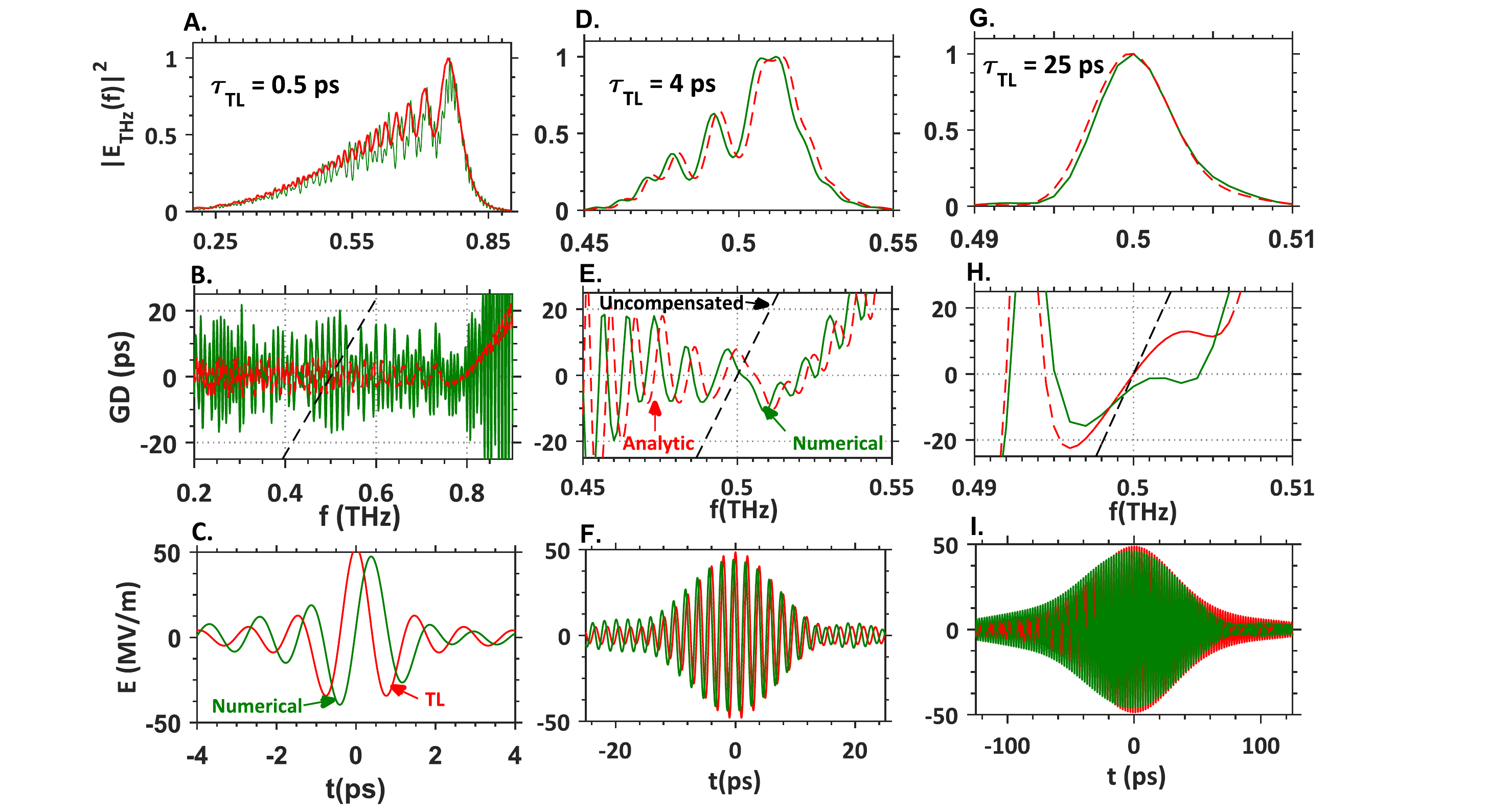}
\caption{\label{fig5}Terahertz spectra and group delay (GD)  obtained from simulations(Eq.\ref{eq:4}, green) and analytic calculations (Eq.\ref{eq:8a}, red-dashed) along with electric field output traces obtained from simulations only (green) and corresponding transform limited values (red-solid). The uncompensated GD due to the relative chirp between pump pulses is also shown (black-dashed). $\tau_{TL}$=0.5 ps (a)-(c), $\tau_{TL}$=4 ps (d)-(f) and $\tau_{TL}$=25 ps (g)-(i). In all cases, the generated pulses are very close to the transform limited output. The ripples in phase and spectra maybe rectified by apodization.}
 
\end{figure*}
In this section, we present the spectra , phase and temporal electric field traces for various values of $\tau_{TL}$ (i.e. pump bandwidths which are proportional to the terahertz bandwidth). In each case, the terahertz frequency was $\text{0.5~THz}$ for optimal values of $\tau\approx 150$ps obtained from previous simulations. 

In Figs.\ref{fig5}(a)-(c), we present the spectrum, phase and temporal format for $\tau_{TL}=\text{0.5~ps}$. The spectral profile in Fig.\ref{fig5}(a) is clearly broadband spanning 0.2 to 0.8 THz. The increasing 'ramp' like shape and ripples are in line with the discussion following Fig.\ref{fig3}. Furthermore, the analytic calculations from Eqs.\ref{eq:8a}-\ref{eq:8c} are in agreement  with numerical simulations based on Eq.\ref{eq:4}. Understandably, the transform limited duration of the corresponding terahertz pulse is proportional to $\tau_{TL}$. For $\tau=0.5$ ps,this corresponds to a single-cycle terahertz pulse as can be seen in Fig.\ref{fig5}(c). 

In Fig.\ref{fig5}(b), the group delay (GD) of the terahertz pulse is plotted as obtained from analysis (Eqs.\ref{eq:8a}-\ref{eq:8c}) as well as simulations (Eq.\ref{eq:4}). In addition, we plot the uncompensated group delay $1/(2b+8\tau^{-4}/b)$ (See derivative of the phase of the first term within brackets of Eq.\ref{eq:7a}) to delineate the extent of phase compensation. As evident from Fig.\ref{fig5}(b), the linear group delay (black-dashed) has been roughly flattened (save for the ripples, also consistent with Eq.\ref{eq:8a}). As a result, in Fig.\ref{fig5}(c), we find that the transform limited trace (red) is very close to the output from the structure. 

Similarly, Figs.\ref{fig5}(d)-(f) and Figs.\ref{fig5}(g)-(h) represent the corresponding properties for $\tau_{TL}$=4 ps and $\tau_{TL}$=25 ps respectively, which both produce pulses (green)close to transform limited values (red).

\subsection{\label{sec:4c}Cascading effects}
\begin{figure}
\includegraphics[scale=0.4]{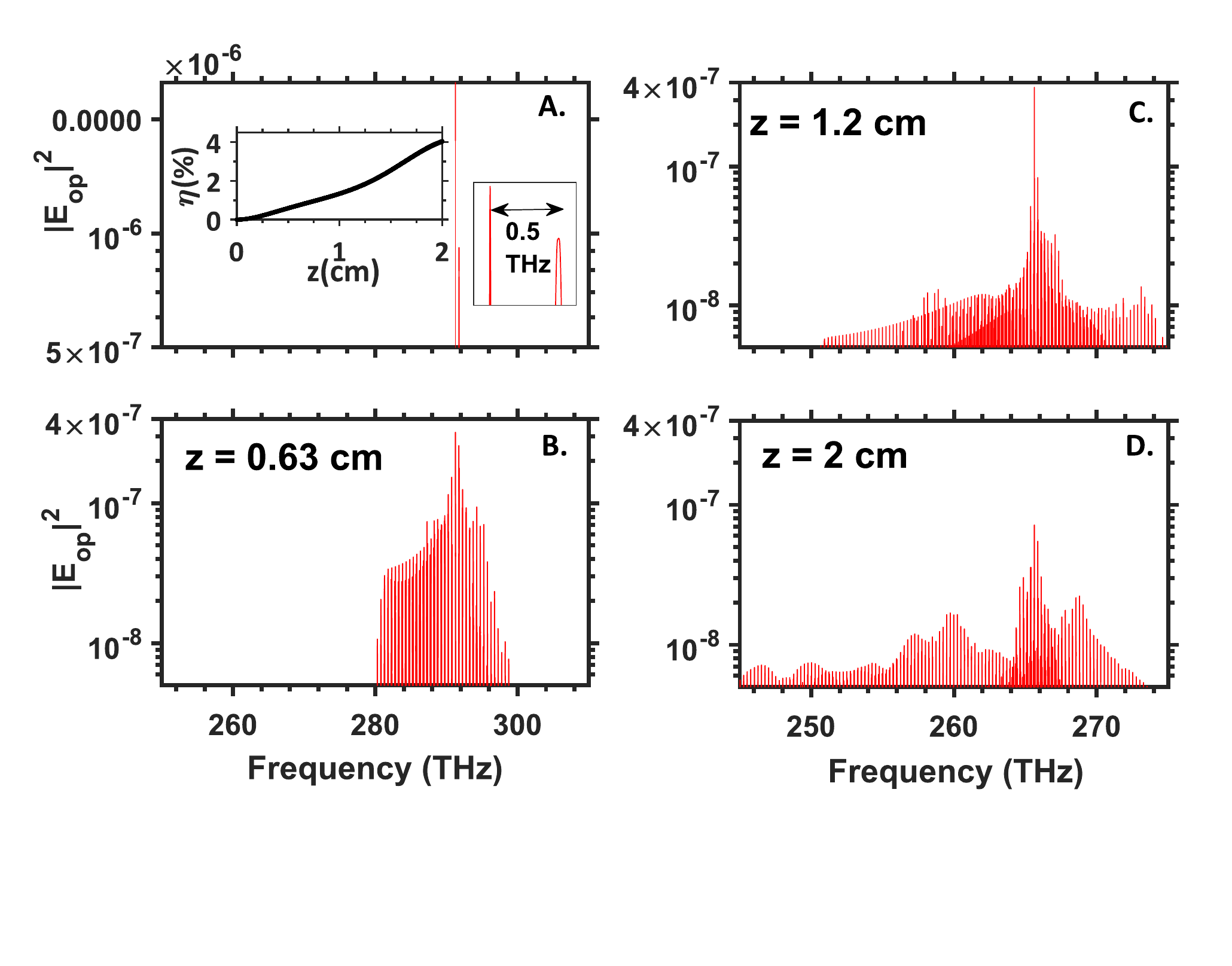}
\caption{\label{fig6} Evolution of the optical pump spectrum as a function of distance. (a) Two 'lines' separated by 0.5 THz, one of which is broader due to a linewidth limited by $\tau_{TL}=$25 ps. The inset on the left shows conversion efficiency approaching 4 $\%$. The efficiency is larger than the corresponding undepleted expression due to the an increase in optical pulse intensity which arises from spectral broadening.(b)-(d)A continuous red-shift is obtained due to the beating of the lines. This red-shift is responsible for the high conversion efficiencies that are observed. Blue shift is also obtained since sum and difference frequency processes are symmetrically phase matched due to the small terahertz frequency} 
\end{figure}

In this section, we resort to simulations in the depleted limit, i.e. including cascading effects based on the numerical model presented in \cite{Ravi:16}. We find that despite drastic spectral broadening of the pump, the efficacy of compression is still maintained and efficiencies are in agreement with calculations in previous sections. 

In Fig.\ref{fig6}(a), the optical spectrum corresponding to two pulses separated by a frequency of 0.5 THz, with parameters $\tau_{TL}=$25 ps, and $\tau=$150 ps is depicted. The choice of these parameters is based on the fact that the expected conversion efficiencies fall in a regime where cascading effects are expected to be significant. For shorter $\tau_{TL}$, the similarity to undepleted calculations would increase owing to lower expectations of conversion efficiency (See Fig.\ref{fig4}(a)).

 Since the energy in the narrowband pulse and broadband chirped pulse are equal, their relative strengths in the spectral domain are dissimilar as is evident via the inset in Fig.\ref{fig6}(a). Note, how  the inset on the right shows differing linewidths. Furthermore, the inset on the left plotting conversion efficiency versus length, shows that the conversion efficiency is similar,albeit higher compared to that obtained via undepleted calculations. The higher efficiencies of $4\%$ is due to the increase in pump intensity resulting from dramatic spectral broadening. 

In Figs.\ref{fig6}(b)-(d), we observe a sifnificant red-shift of the optical spectrum due to cascading effects, which are responsible for the high efficiency in the first place. Since, the strength in the two input lines differ, the situation is similar to that of a cascaded terahertz parametric amplifier \cite{raviCOPA:16}. The spectrum shows an increasing red-shift with distance, accompanied also with sum frequency generation due to symmetry in phase-matching of sum and difference frequency generation that results from the small value of the terahertz frequency, similar to that described in\cite{raviCOPA:16}.

\begin{figure}
\includegraphics[scale=0.23]{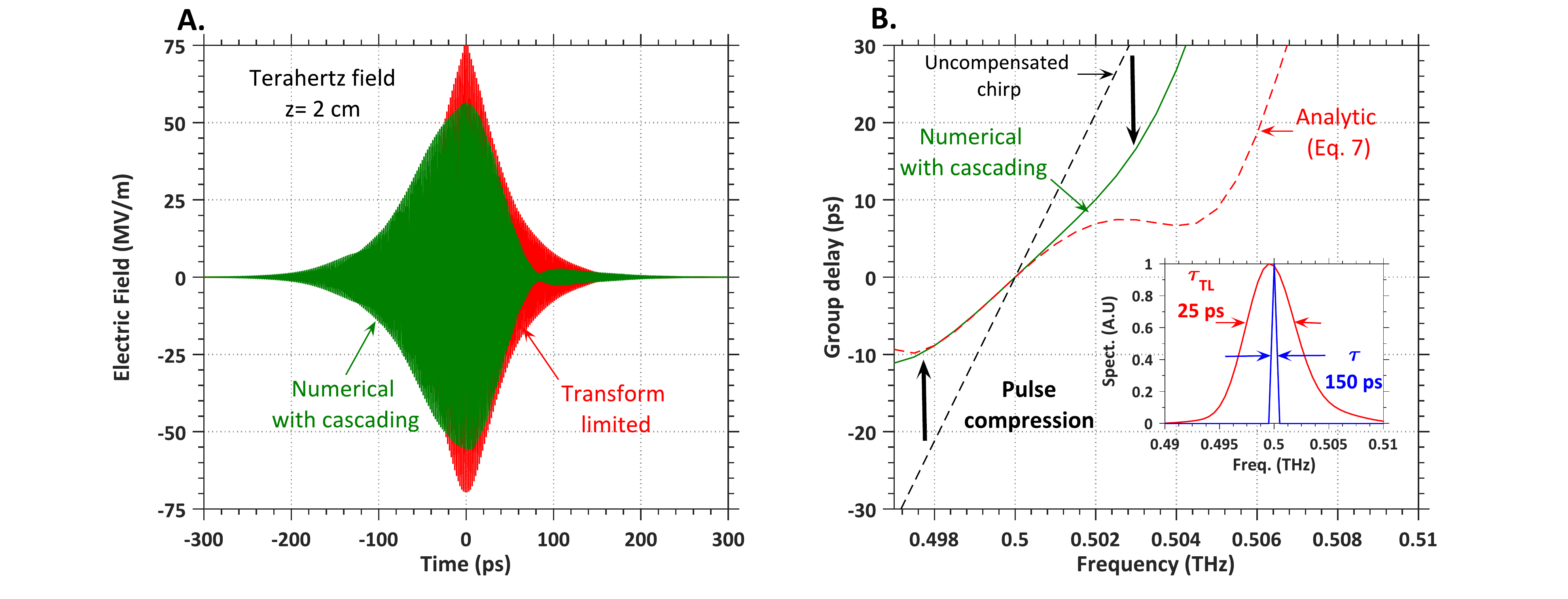}
\caption{\label{fig7} (a) Traces of the terahertz electric field obtained from simulations accounting for cascading (green) and the corresponding transform limited terahertz pulse(red). The output from the structure is similar to the transform limited value and has a peak field strength similar to that predicted from undepleted calculations. (b) Group delay of compressed pulses obtained from simulations (green) and undepleted analysis (red-dashed) show effects of phase compensation in relation to the chirp present in the pump input (black-dashed). The analytic calculations show superior compression for $>.502$ THz, due to the effects of dispersion arising from cascading. (Inset) Most of the spectrum lies in a range where the undepleted and depleted calculations are in agreement, which results in effective compression. The spectrum of a narrowband pulse corresponding to $\tau=150$ps(inset,blue) is shown for comparison. As efficiencies reduce, cascading effects would be less important and the efficacy of compression would approach that predicted by undepleted calculations. }
\end{figure}

In Fig.\ref{fig7}(a), we plot the simulated terahertz field output and compare it to the transform limited terahertz field (obtained by assuming a flat phase across the simulated spectrum). The peak field of the transform limited field is about $75$ MV/m (slightly larger than the TL value of 60 MV/m in Fig.\ref{fig5}(I) due to the increase in terahertz efficiency by cascading effects), while the output from the structure is close but smaller at $55$ MV/m (similar to peak fields obtained from undepleted calculations in Fig.\ref{fig5}(I)).  The durations of the two pulses are quite similar, indicating effective compression. 

Figure \ref{fig7}(b), depicts the group delay obtained numerically while considering cascading effects (green) and analytically under undepleted approximations(Eq.\ref{eq:8a},red-dashed) in comparison to the uncompensated chirp of the pump pulse input (black-dashed). Clearly, the chirp on the pump input has been significantly reduced by the structure as evident in both simulations and analytic calculations. However, for frequencies $>0.502$THz, the undepleted calculations suggest better compression. However, since most of the energy in the spectrum is contained within the region where undepleted and depleted calculations agree (see inset of Fig.\ref{fig7}(b)), effective compression is nevertheless obtained. To delineate how broadband the generated terahertz spectrum is, we also plot the spectrum of a narrowband pulse corresponding to $\tau=150$ ps in the inset.

The difference between depleted and undepleted calculations is due to the change in optical group refractive index encountered by the broadened spectrum. This renders the generated poling profile less effective. Since, the extent of broadening is efficiency dependent, the disparity between depleted and undepleted calculations can be expected to reduce for even more broadband terahertz generation (e.g. Fig.\ref{fig5}(a)), where the efficiencies are anticipated to be lower.

In the high efficiency regime delineated in Figs.\ref{fig6} and \ref{fig7}, the compression maybe further improved by numerical optimization of the poling profile, which would be the scope of future work.

However, it is worthwhile to note that the  poling profile conceived analytically in the absence of depletion is highly effective, even in the presence of cascading effects producing conversion efficiencies $>4\%$.

\section{\label{sec:5}Conclusion} 
In conclusion,a new method to generate broadband terahertz pulses using a combination of chirped pulses and aperiodically chirped crystals was shown to produce compressed pulses with peak fields and high efficiencies. The approach is particularly efficacious for generating few to tens of cycles of pulses, yielding conversion efficiencies of $\approx 5\%$. Analysis in the undepleted limit along with simulations in undepleted and depleted limits were presented. The simulations validated the efficacy of the approach. In the presence of dramatic cascading, the efficacy of compression only reduces marginally but this can be further improved by numerical optimization of the poling profile. The presented work could be potentially disruptive for high field terahertz acceleration and coherent X-ray generation.  


\section{Acknowledgments}

This work was supported under by the Air Force Office of Scientific Research under grant AFOSR - A9550-12-1-0499
the European Research Council under the European Union's Seventh Framework Programme (FP/2007-2013) / ERC Grant Agreement n. 609920 and the Center for Free-Electron Laser Science at DESY.

\bibliography{apssamp}

\end{document}